# Self-Calibrated Epipolar Reconstruction for Assessment of Aneurysms in the Internal Carotid Artery Using In-Silico Biplane Angiograms

**Kyle A. Williams,[a,d] SV Setlur Nagesh,[c,d] Daniel R. Bednarek,[a,b,d] Stephen Rudin,[a,b,c,d] Ciprian N. Ionita[a,b,c,d,*]**

[a]University at Buffalo, Department of Biomedical Engineering, Buffalo, USA

[b]University at Buffalo, Department of Radiology, Buffalo, USA

[c]University at Buffalo, Department of Neurosurgery, Buffalo, USA

[d]Canon Stroke and Vascular Research Center, Buffalo, USA

*Ciprian Ionita, E-mail: cnionita@buffalo.edu

**Abstract**

**Background:** The treatment of intracranial aneurysms (IA) predominantly relies on angiography guidance using biplane views. However, accurate flow estimation and device sizing for treatment are often compromised by vessel overlap and foreshortening, which can obscure critical details.

**Purpose:** This study introduces an epipolar reconstruction approach to enhance 3D rendering of the internal carotid artery (ICA) and aneurysm dome using standard, routinely acquired biplane angiographic data. Our method aims to improve procedural guidance and device selection by overcoming the limitations of traditional two-dimensional imaging techniques.

**Methods:** This study employed three 3D geometries of ICA aneurysms to simulate virtual angiograms, including the aneurysm dome, parent vessel, tortuous carotid cavernous segment and the ICA terminus bifurcation. Virtual angiograms were generated using a computational fluid dynamics (CFD) solver, followed by the simulation of biplane angiography using a cone-beam geometry. Self-calibration was accomplished by matching contrast media position as a function of time between biplane views. Feature-matching based on axial positioning was used to triangulate and reconstruct the vascular structures in 3D. The projection data was utilized to refine the 3D estimation, including elimination of erroneous structures and ellipse-fitting. The accuracy of the reconstructed images was quantitatively evaluated using the Dice-Sorensen coefficient, comparing the 3D reconstructions to the original CFD-generated models to assess the geometric fidelity.

**Results:** The proposed epipolar reconstruction method generalized well across the three tested aneurysm models, with respective Dice-Sorensen coefficients of 0.745, 0.759, and 0.654. Errors were primarily due to partial vessel overlap, observed in our third model. The average reconstruction time for all three volumes was approximately 10 seconds.

**Conclusions:** The implemented epipolar reconstruction method enhanced 3D visualizations from biplane angiographic data, addressing key challenges such as projection-induced vessel foreshortening. This method provides a solution to the complexity of IA visualization, with the potential to provide more accurate analysis and device sizing for treatment.

# 1   Introduction

A major challenge in the treatment of intracranial aneurysms (IAs) is the reliance on 2D imaging during neurointerventional procedures, which often fails to fully capture the complex three-dimensional (3D) anatomy of the vascular structures. Standard biplane angiography, the most commonly used imaging modality in neurovascular procedures, provides high-resolution 2D projections but is limited by vessel overlap and foreshortening, particularly in tortuous and bifurcated regions such as the internal carotid artery (ICA). These limitations hinder accurate visualization and quantitative analysis, complicating device selection and placement during interventions.

Although cone-beam computed tomography (CBCT) offers detailed three-dimensional insights and could be utilized for procedural guidance, its application is often impractical during interventions[1,2]. Limitations include the availability of CBCT in certain clinical settings and its inability to provide real-time imaging updates during the navigation and deployment of devices. As a result, biplane angiography remains the standard imaging modality in these procedures, despite its inherent two-dimensional limitations.

One approach to mitigate the 2D imaging restriction in angiographic systems is epipolar reconstruction from biplane angiography. This approach compares two projection images of the same object and estimates the epipolar geometry of the system to reconstruct a 3D representation of the object (Figure 1). Our research has identified gaps in the application of epipolar geometry for automatic vascular reconstruction in settings characterized by high vascular tortuosity, aneurysms, and bifurcations—common features in the ICA. Previous studies have explored epipolar reconstruction but have not adequately addressed these specific challenges in an automated way[3-6], or without the use of a calibration object.[7,8]

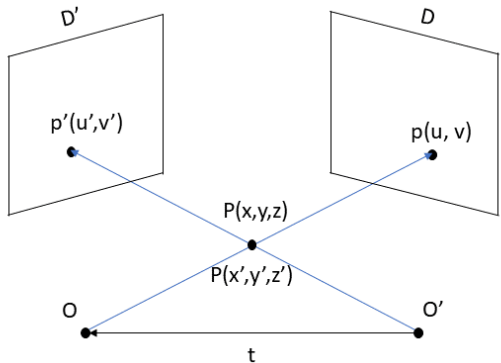

**Figure 1:** Geometry of a biplane transmission imaging system used for epipolar reconstruction. Unprimed coordinates are relative to the first imaging system including observer, O (the x-ray source), image plane, D, and projection p of a point P in the object, while primed coordinates are relative to the second imaging system including O', D' and p'.

In this study, we address these challenges by proposing a novel, self-calibrated epipolar reconstruction technique[9-11] that leverages routinely acquired biplane angiographic data to generate high-fidelity 3D reconstructions. Our approach is specifically designed to overcome the bottlenecks posed by vessel tortuosity, aneurysm morphology, and bifurcations in the ICA. By combining feature-matching algorithms with advanced projection filtering, the method eliminates the need for external calibration objects and ensures robustness in complex anatomical scenarios.

## 2 Materials and Methods

### 2.1 In-silico angiogram generation

CTA-derived patient-specific vascular phantoms were imported into ICEM (ANSYS Inc., Canonsburg, PA), a CFD mesh generation software. Inlet and outlet boundaries were identified for each model, and simulation parameters consistent with previously reported CFD-based simulated 4D angiographic sequences were used[12]. We performed transient, laminar flow simulations by numerically solving the incompressible Navier-Stokes equations in ANSYS Fluent (ANSYS Inc., Canonsburg, PA), assuming rigid walls and a Newtonian fluid with a density of 1060 kg/m$^3$ and viscosity of 0.0035 Pa-s. The time step was set at 1 ms to provide sufficient temporal resolution to resolve all unsteady motions in the flow. This ensures the convergence criteria of 1e-6 was achieved at each time step and provides a direct comparison to *in vitro* angiography. Using the passive scalar method[13], fluid was labeled as a tracer at the arterial inlet over a 1.0-second interval and allowed to propagate through the vessel, where tracer concentration is distributed within the range [0, 1] per voxel. The simulation continued for 500 time steps (0.5 s) after tracer labeling was disabled for a total acquisition time of 1.5 seconds, allowing sufficient time for simulated contrast to fully exit the vascular geometry. This process results in a 4D simulated angiogram with a 1.0-second contrast injection and sufficient temporal coverage to capture both the inflow and outflow of contrast within the vessel of interest. This process was repeated for three intracranial aneurysm models, resulting in three 4D angiography simulations.

To transform these 3D simulations into two-dimensional biplane angiograms, a cone beam geometry was employed in ASTRA to simulate projection imaging with realistic geometry and variable magnification factors. By fully simulating the image formation, we can precisely control and understand the relationships between 2D and 3D representations, allowing more accurate comparison of ground truth and reconstructed data.

## 2.2  Image Pre-processing for an Uncalibrated System

Following the generation of biplane angiographic views through cone-beam geometry, the subsequent challenge lies in effectively processing these images for 3D reconstruction without calibration to standardize the sizes of objects between both images. Traditional epipolar reconstruction methods often falter when applied to uncalibrated medical imaging data, particularly due to difficulties in accurately determining pixel size scaling factors and FOV alignment to represent structures of interest in a geometrically consistent way[14]. An important step in calibration between imagers is a standardization of the pixel size in each imager, as a function of pixel pitch and magnification factor. While it is not possible to determine absolutely pixel size without a calibration object of known size, it is possible to ensure that the pixel size of one image is the same as the other by estimating the magnification of the vessel in each image.

We propose a fully automatic, self-calibration method to determine the scaling factor between images using the transit distance of contrast media between subsequent frames, which is common for both views in the axial direction. Since biplane imaging systems share one common axis between the two imagers (typically in the axial direction), it is possible to determine an equality in the system in which, over an interval of time, the advancement in contrast agent observed in the angiographic data should be the same in both imagers, shown in Figure 2. Scaled by the known pixel pitch of each detector, contrast advancement from correspondent time points is used as a surrogate to true knowledge of the magnification factors of each imager. The primed system is rescaled to the same magnification as the reference unprimed system, effectively applying the scaling factor, k, which is the ratio of contrast bolus transit distances (l and l') between the images for the two projections. This is similar in concept to an algorithm proposed to reduce ghosting artifacts in epipolar reconstruction, which utilized both bolus arrival time and centerline velocities as constraints on the vascular tree[6], however, we propose its use as an initial constraint for magnification factor equalization. Here, we utilize only the bolus arrival time due to its efficiency, robustness to quantum mottle, and independence from vessel centerline determination.

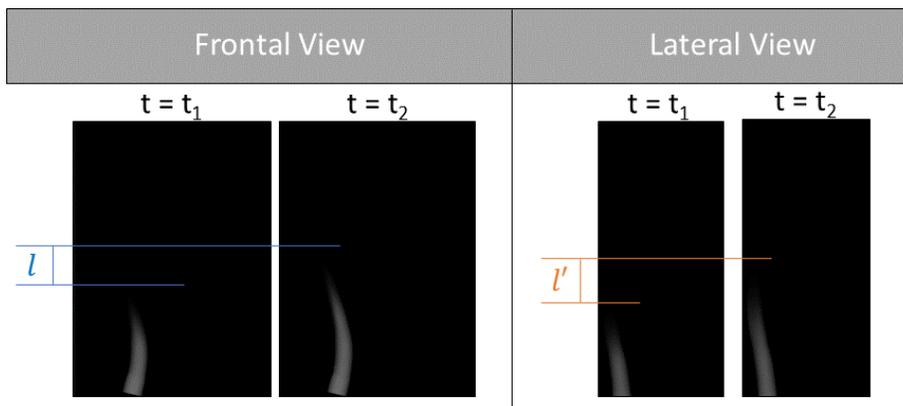

**Figure 2:** Time points $t_1$ and $t_2$ are compared in both frontal and lateral views after pixel pitch equalization. The advancement of contrast in each view is measured by taking the difference (in pixels) between the top-most location of contrast media in $t_2$ and $t_1$, denoted $l$ and $l'$ for the frontal and lateral measurements, respectively. The ratio of these values is used to equalize the magnification factor between frontal and lateral projections.

A similar paradigm is used to accomplish axial alignment of biplane views. Here, we utilize the distance between the highest axial point of contrast and the bottom of the image frame at corresponding time steps in each image system, shown in Figure 3. Since single pixels represent the same physical distance in each imager after rescaling, the only remaining discrepancy between the two systems is the field of view (FOV). To ensure all axial points in the image have a match, whichever imager has a smaller FOV is used as the reference, and the other image is cropped accordingly.

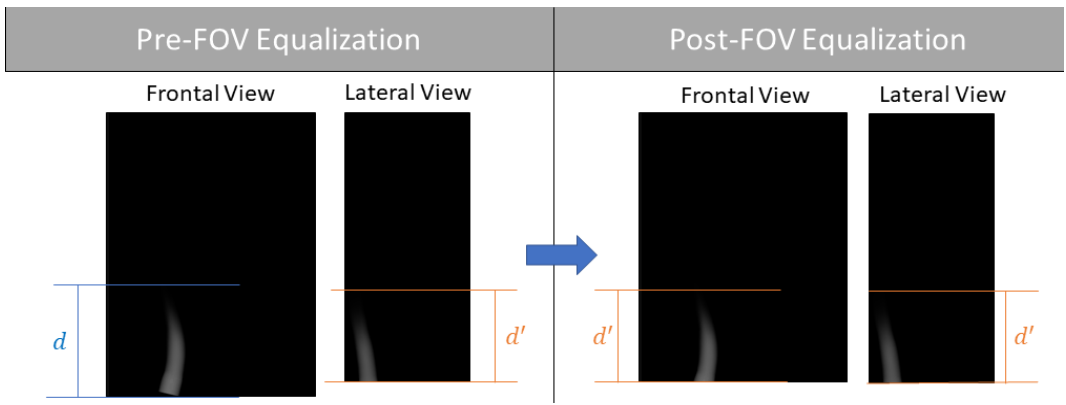

**Figure 3:** After magnification factor equalization, representation of contrast advancement (in pixels) should be equivalent between views. For a given time step, any remaining discrepancies in the distance from the bottom of the image to the top-most point of contrast media between views is attributed to FOV misalignment. This measurement is used to compare and standardize the images, ensuring the FOV of both imagers is consistent, and all axial slices will be aligned between views.

## 2.3 Epipolar Reconstruction

Once the images are scaled and aligned, feature matching is accomplished by comparing the axial positions of objects in each image, possible due to the previous alignment step. The matched features between the imaging systems are used as input points to the essential matrix solver available through Python's implementation of the OpenCV library[15]. The library has additional functions that determine the correct rotation and translation matrices, providing an estimate of the rotation angle between the imagers.

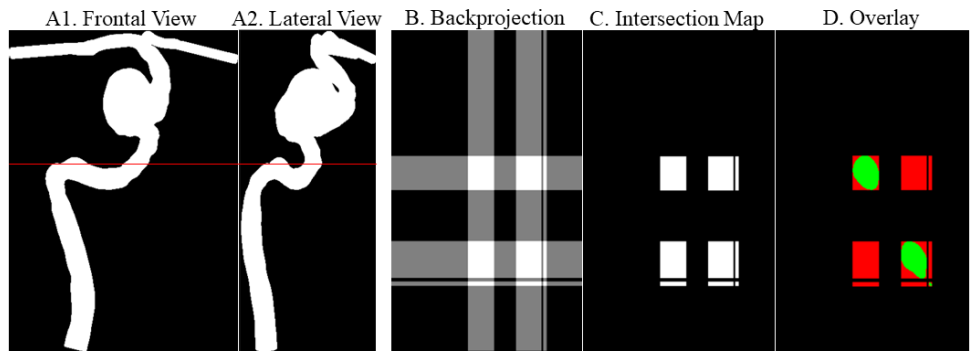

**Figure 4:** Binary vascular masks from each projection view (A1, A2) are back-projected into a common volume. A single axial slice of this volume (B), the position of which is indicated with red lines in A1 and A2, shows the contribution of the frontal view (projected from left to right) and the lateral view (projected from top to bottom). This back-projection map is used to identify intersections between views (C) across all axial slices, though extraneous structures must still be filtered out (D). Here, green indicates true positive, and red indicates false positive.

This angulation, along with the dimensions of the images from each view, is used to back-project the vessel masks from each view into a common volume. The raw output of this step is shown in Figure 4B. Taking the intersection of the back-projections from each view, we observe that the output is the smallest bounding box which fully encompasses the vessel at each axial slice, as shown in Figure 4C. While this approximation yields accurate results for singular vascular structures, artifacts caused by multiple structures in a given slice still need to be addressed. In axial slices where multiple objects are detected in each view, the algorithm cannot effectively pinpoint which bounding boxes contain true vascular structures, as observed in Figure 4D, where the ground truth structure is overlaid onto the epipolar algorithm's raw output.

To eliminate erroneous structures in each slice, we compare each possible combination of bounding boxes to a temporal maximum intensity projection (temporal MIP) of the biplane angiography data. Assuming a uniform distribution of contrast, we estimate the thickness of the structure or structures present in each slice of the biplane data to be the maximum intensity within the structure and assign this value to the structure. This is repeated for both projection views to determine an estimated length and width of the target bounding boxes, yielding constraints to both bounding box size, and position. If the estimated number of structures existing in a particular slice is too many or too few, or if the estimated structure is much larger than the originally projected object, the structure (or combination of structures) will be eliminated as a potential solution. This process can then be repeated at each axial slice, where only the best match is taken per slice. Though this will substantially reduce the number of possible objects in each slice, it can result in a discontinuous 3D reconstruction, as alternative erroneous solutions may potentially also register as best matches, varying by axial slice.

Instead, we employ structural projection filtering, where axially continuous structures are automatically identified and individually labeled as objects throughout the corresponding axial slices. The same "best match" approach is used per combination of continuous structures (per axial slice), with the combination of structures most closely matching the original projection data across all slices chosen as the correct solution. This approach allows more efficient combinatorial testing as there are fewer testable combinations if axial continuity is used as a constraint and ensures that the resultant 3D structure is continuous. Figure 5 shows several axial slices before and after structural projection filtering.

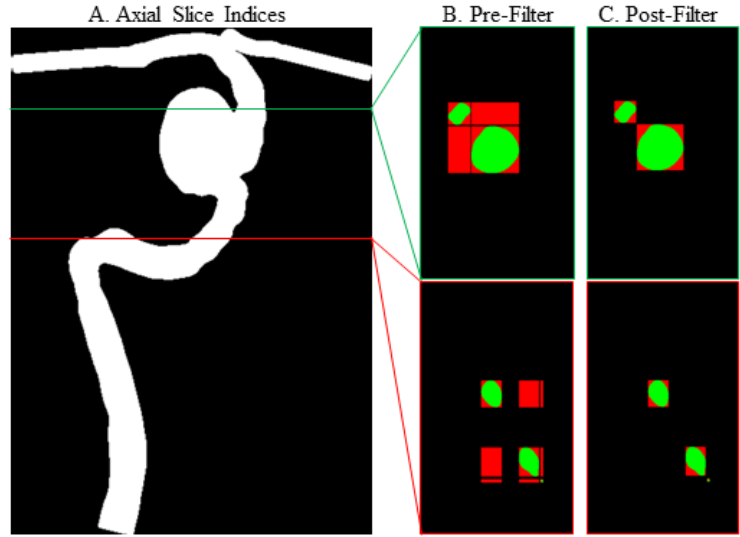

**Figure 5:** Several axial slices (A) of the reconstructed volume before (B) and after (C) structural projection filtering. We observe the correct elimination of bounding boxes where several erroneous positions were originally recorded.

### 2.4 Ellipse-Fitting Refinement

Once structural projection filtering is completed, the vessel of interest should exist entirely within the remaining bounding boxes, as shown in Figure 6. Although this gives a good estimate of the maximal length, maximum width, and position of the vessel in each axial slice, it poorly represents the shape of the vasculature. Instead, we assume that an ellipse may approximate the cross-section of most vascular structures of interest well[16,17].

However, if the Cartesian axes are simply used to calculate the shape of the ellipse, eccentric objects are poorly represented. Instead, we again utilize our projection data. Since projection imaging already integrates each shape in the depth direction, we can define the line profile of the projected object as a single integral of the object in 3D space (Eq. 1).

$$C(x,z) = \int_0^{y_{max}} C(x,y,z)dy \qquad (1)$$

where C(x, z) is the density of iodinated contrast along the line profile of an object as a function of its in-plane position, x, and axial position, z, C(x,y,z) is the 3D distribution of contrast densities, and depth-integration occurs along the y-axis (a similar relation exists for C(y, z) in the second biplane view). If we take an integral of this line profile (Eq. 2), it represents a double integration of an axial cross-section of the vascular structure, which can be used to approximate the cross-sectional area of the vessel.

$$\int_0^{x_{max}} C(x,z)dx = \int_0^{x_{max}} \int_0^{y_{max}} C(x,y,z)dy\,dx \cong Area(z) \tag{2}$$

where Area(z) is the cross-sectional area of the object as a function of its axial position, z. It's worth noting this calculation can be performed from either view with similar results, provided the differences between pure integration and exponential attenuation of x-rays are factored correctly.

With our given information, we are tasked with generating an ellipse that exists entirely within, and extends to the full height and width of, the given bounding box and closely resembles the object area, calculated via line profile integration. We first start with the angled ellipse equation (Eq. 3).

$$\frac{(xcos\theta - ysin\theta)^2}{a^2} + \frac{(xsin\theta + ycos\theta)^2}{b^2} = 1 \tag{3}$$

where x and y are the coordinates of the ellipse, θ is the ellipse angle, and a and b are the semi-major and semi-minor axes, respectively. Taking its derivative, we can calculate the instantaneous slope of the ellipse as a function of its x- and y-coordinates (Eq. 4).

$$\frac{dy}{dx} = \frac{a^2\,x\sin^2\theta + y(a-b)(a+b)sin\theta cos\theta + b^2 x cos^2\theta}{a^2 y cos^2\theta + x(a-b)(a+b)sin\theta cos\theta + b^2 y sin^2\theta} \tag{4}$$

If we set the numerator to zero, we find the horizontal tangents to the ellipse, indicating the location of our upper and lower bounds of the y-coordinates of the bounding box encompassing this generalized ellipse. If we set the denominator to zero (i.e. undefined slope), we find the vertical tangents to the ellipse, indicating the upper and lower bounds of the x-coordinates of the bounding box. This allows us to set up a system of equations and solve for the x and y bounds of the ellipse (Eqs. 5 – 6).

$$a^2\,x\sin^2\theta + y(a-b)(a+b)sin\theta cos\theta + b^2 x cos^2\theta = 0 \tag{5}$$
$$a^2 y cos^2\theta + x(a-b)(a+b)sin\theta cos\theta + b^2 y sin^2\theta = 0$$

Solving for x and y we obtain the span of the bounding box relative to its centroid,

$$x = \pm\sqrt{a^2 \cos^2\theta + b^2 \sin^2\theta}, \;\; y = \pm\sqrt{a^2 \sin^2\theta + b^2 \cos^2\theta} \tag{6}$$

However, for our application, we have a fixed bounding box and need to determine the lengths of the ellipse's semi-major and semi-minor axes as a function of the bounding box size and angle theta (Eq. 7).

$$a = \pm\sqrt{x^2 - \frac{\frac{y^2 \cos^2\theta - x^2 \sin^2\theta}{\cos^2\theta - \sin^2\theta}\sin^2\theta}{\cos^2\theta}}, \quad b = \pm\sqrt{\frac{y^2 \cos^2\theta - x^2 \sin^2\theta}{\cos^2\theta - \sin^2\theta}} \qquad (7)$$

Finally, with a and b determined, the area of the ellipse can be calculated using Eq. 8.

$$Area = \pi a b \qquad (8)$$

With these relationships in mind, for a given bounding box and estimated cross-sectional area, we determine the area of a set of ellipses that span the bounding box as a function of their orientation, θ, and choose the orientation that best matches the target cross-sectional area. This process is illustrated in Figure 6. By fitting an ellipse parametrically (as a function of its semi-major and semi-minor axis lengths and its orientation, θ), the algorithm removes the need for iterative refinement, substantially improving its efficiency.

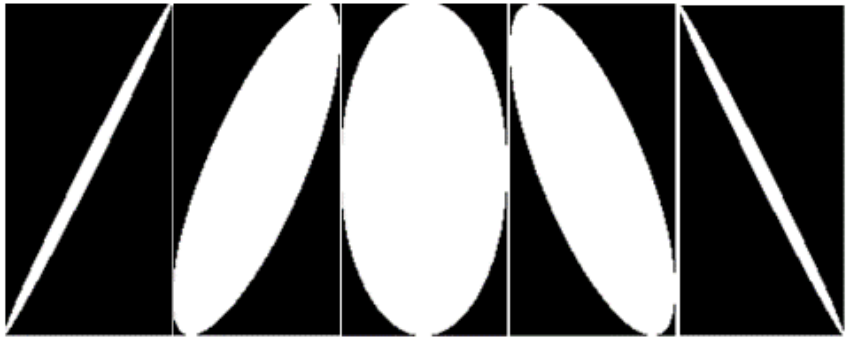

**Figure 6:** Many ellipses with varying orientations are possible for a given bounding box. To remain within the span of the bounding box, the major and minor axes of the ellipse must be modulated, leading to differences in their total areas.

Since each generated ellipse is forced to pass through the centroid of the bounding box, it is symmetric about its Cartesian axes. This means that, for angles ranging [0, 360], eight different ellipse orientations give the same area (object symmetry reduces this to only two orientations). To choose a single correct orientation, the orientation from the previous axial slice of the same object is stored, and the angle which most closely matches that of the previous slice is chosen as the singular correct orientation. The output of this step is a well-refined approximation of the true vascular structure, shown in Figure 7. To quantitatively determine the alignment between this refined 3D estimate and ground truth 3D structure from CFD, the Dice-Sorensen coefficient between the two datasets was calculated across the entire volume and reported.

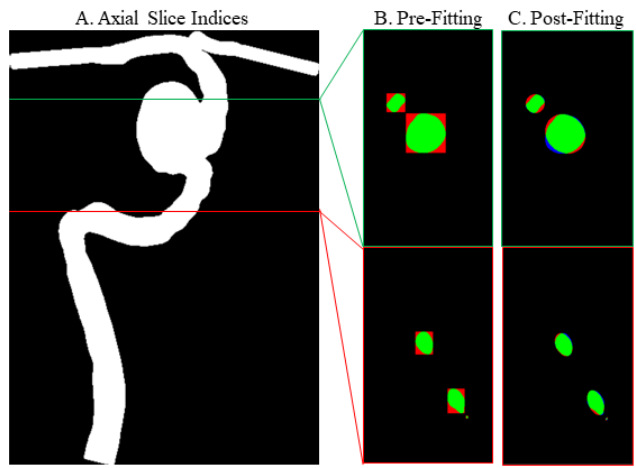

**Figure 7:** Examples of ellipse fitting at two different axial slices of the vessel of interest. Here, green indicates true positive, red indicates false positive, and blue indicates false negative.

## 3    Results

### 3.1    Reconstruction of 3D Vascular Structures from Two Views

The reconstruction algorithm performed consistently across each of our three test models, shown in Figure 8, with an average reconstruction time of approximately 10 seconds. The Dice-Sorensen coefficient for models M1, M2, and M3 were 0.745, 0.759, and 0.654, respectively. Errors primarily arose in regions with severe vessel overlap, where intensity-based refinement techniques, such as structural projection filtering and ellipse fitting struggled to separate the intensity profiles of overlapped objects. A comparison of reconstructions for axial slices with and without overlapping structures is shown in Figure 9.

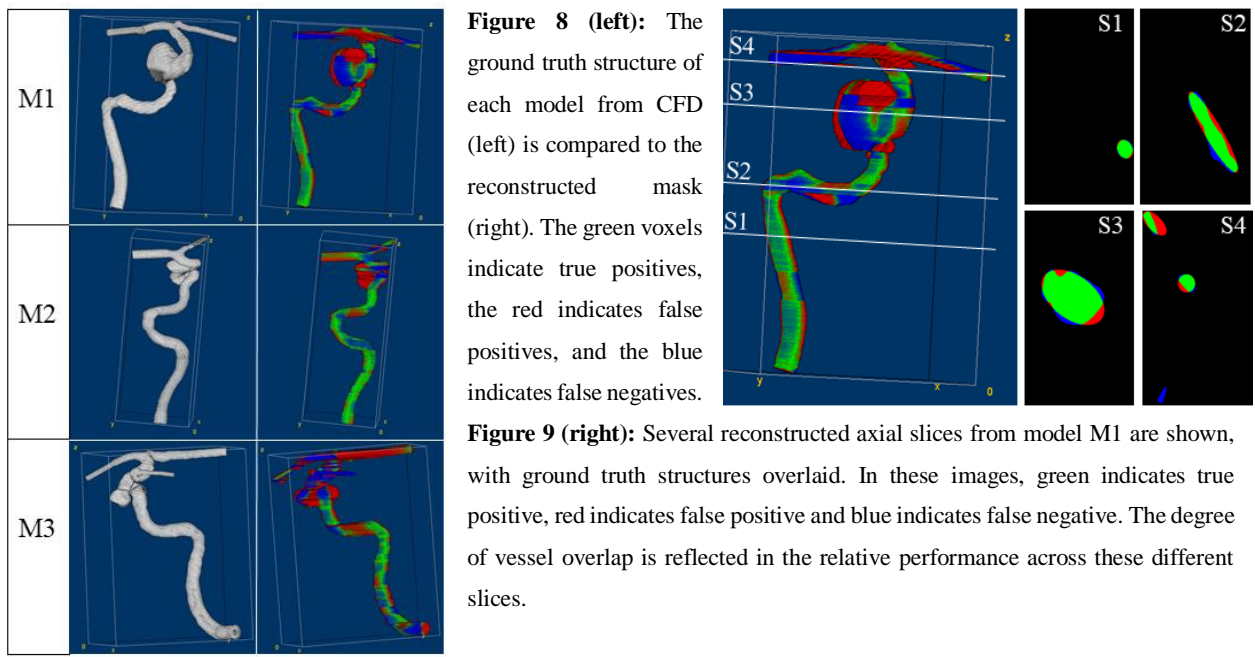

**Figure 8 (left):** The ground truth structure of each model from CFD (left) is compared to the reconstructed mask (right). The green voxels indicate true positives, the red indicates false positives, and the blue indicates false negatives.

**Figure 9 (right):** Several reconstructed axial slices from model M1 are shown, with ground truth structures overlaid. In these images, green indicates true positive, red indicates false positive and blue indicates false negative. The degree of vessel overlap is reflected in the relative performance across these different slices.

## 4    Discussion

The reconstruction algorithm discussed above presents many advantages when applied to clinical scenarios. First, the alignment method described in Section 2.2 purely utilizes properties of the imaging data itself. This means that, with inaccurate or absent DICOM headers regarding magnification, pixel pitch or FOV, and in the absence of a calibration object, image-based comparison is still robust. While the method is generally effective, it is worth noting that its accuracy depends on temporal correlation between biplane views, such that both views are taken as close to simultaneously as possible. Though this correlation is strong at higher frame rates such as the data observed in this study as well as several in vitro and in vivo investigations utilizing 1000 fps high-speed angiography[12,18-23], the alignment technique may prove difficult for lower frame rates, such as the 3 fps sequences commonly acquired in the clinic. This could be mitigated via iterative estimation of the magnification factors in each view[24], or simultaneous acquisition of both views, though this introduces tradeoff in accuracy for projection-based filtering and ellipse fitting techniques due to increased cross-talk scatter between views. It should also be noted that the physical correlation in axial travel distance of contrast media between two views relies on each view sharing the same axis in the axial direction. This means cranial or caudal angulation in either gantry may introduce bias in the magnification or FOV equalization steps of pre-processing the data, though this could be corrected using more advanced image processing techniques, such as affine transformations[25].

The second advantage of this algorithm is the ability to render vascular structures of interest from biplane angiography, with no additional imaging requirements. This not only presents potential dose savings to patients, where the need for one or more CTA acquisitions throughout the course of their treatment may be obviated, but also reduces the need to transfer patients from one imaging suite to the next, improving the net efficiency of the clinic and the quality of care. Additionally, since flat panel detectors have isotropic pixel sizes with high resolution, it becomes very efficient to generate high-fidelity, isotropic voxel representations of the vessel of interest using this approach.

It should be noted that, since the reconstruction algorithm is both non-iterative, and rule-based, the 3D reconstruction approximates the true structure of the vessel and is subject to error in vascular regions with partial overlap. This overlap-based error presents itself as an over approximation of the bounding box length, width, or both (depending on which vessels overlap each other in each view), shown in Figure 10. This may restrict the usage of the current algorithm to vasculature which is clearly visible in both biplane views, however, further refinement, either with iterative methods or neural network-based approaches, may yield improved approximations of the vascular network. These methods were not explored during this study, as the most direct, efficient, and easily explainable method of generating 3D reconstructions was preferred.

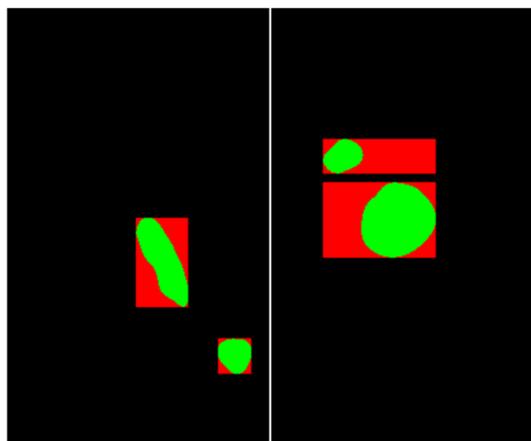

**Figure 10:** Bounding box generation in two axial slices of model M1. In these images, green indicates true positive, and red indicates false positive. In the left image, there is no overlap between vascular structures in either view, allowing clear localization and definition of vessel boundaries. However, in the right image, we observe an overlap in the lateral profile (projected from the top to the bottom), leading to an extension of the bounding boxes of both vessel structures from left to right.

Another potential source of error in the structural refinement or ellipse fitting steps is the incomplete filling of the entire vasculature with contrast. Although this is not a problem for our simulated data, which includes a large, continuous bolus of contrast, it may present as an underapproximation of the structure of the vessel should contrast not reach peak opacity. Fortunately, this phenomenon can be mitigated by taking a temporal MIP of the angiogram as described in Sect. 2.3. This increases the probability that, despite a lack of a singular time point where the vessel is fully filled, the vessel is represented at peak opacity across its volume.

There are factors specific to real angiographic images that must be addressed before applying this technique to such data. First, our simulated angiograms are free of quantum mottle, which introduces variability in intensity values across the image. While this may not significantly impact the creation of vessel masks from each biplane view, it could affect algorithmic steps that rely on intensity profiles for refining the 3D vascular structure, such as structural projection filtering or ellipse fitting. Additionally, in our simulations, the biplane data represent a direct depth integration of contrast-filled vessels, resulting in contrast intensities equivalent to the integrated thickness of the vessel along the projection axis. For real angiographic data, however, it is essential to account for x-ray attenuation. A conversion factor must be applied, or a log-corrected algorithm should be used, to relate the raw intensity values to voxel-based thickness. As demonstrated in previous studies,[26] this can be achieved by inversely solving the exponential attenuation equation for the x-ray path length through the vessel at each location. This could be approximated by inversely solving the exponential attenuation equation (Eq. 12) for x at each location in the vessel

$$\frac{I}{I_0} = e^{-\mu x}, \qquad x = \frac{\ln\left(\frac{I_0}{I}\right)}{\mu} \tag{12}$$

where $I$ is the final x-ray intensity (recorded on the image plane), $I_0$ is the initial X-ray intensity (prior to passing through the vessel), $\mu$ is the linear attenuation coefficient of iodine, and x is the thickness of the vessel through which the x-ray has passed. Information to determine intensity and beam energy should be available from DICOM header information, while effective iodine attenuation coefficients could be provided in a look-up-table. Building on these capabilities, the proposed algorithm holds significant potential for future clinical applications in the treatment of IAs within the ICA. The method could enhance intraoperative navigation and device placement by offering improved spatial visualization of the relationship between devices and vascular structures. Additionally, the approach could be further developed into a correction algorithm for 2D angiography to minimize view bias, ensuring more accurate quantitative assessments across different imaging perspectives. These advancements would enhance the visualization of the neurovascular structure and improve depth-axis context for better quantitative angiographic accuracy[27-30].

## 5    Conclusions

This study presents a novel epipolar reconstruction technique to determine the 3D morphology of vascular structures of interest, automatically and without the use of a calibration object. The technique could prove useful in neurointerventional suites, both to obviate the need for repeat CTA protocols, and to reduce the need for patient transfer between imaging suites, provided complications of real angiographic data, such as exponential attenuation of x-rays and variance in attenuation as a function of polychromatic beam energy, are carefully addressed. Implementation of the method could reduce radiation dose to patients in the neurointerventional suite and improve treatment efficacy, contributing to improvement of patient outcomes. Additionally, determination of 3D vascular structure could enhance the accuracy of a number of quantitative angiography algorithms, further improving treatment specificity and standard of care in neurovascular interventions.

**Acknowledgements**

This work was supported by NSF STTR Award # 2111865.

**Conflict of Interest Statement**

The authors of this manuscript have no conflicts of interests to disclose.